\documentclass[final,5p,times,twocolumn]{elsarticle}

 \usepackage{graphics}

 \usepackage{epsfig}

\usepackage{amssymb}

\begin{document}

\begin{frontmatter}

\title{\bf THE ENERGY PROBLEM }

\author{Giorgio Giacomelli\\ 
Physics Department of the University of Bologna and INFN Sez. of Bologna ~~{\footnotesize (giacomelli@bo.infn.it)}\\
\vspace{0.3cm}
{\large Special general lecture given at the 24$^{th}$ ICNTS Conference, 1-5 September 2008, Bologna, Italy.}}

\begin{abstract}

 A brief general picture of the world, european and italian energy situations is made, analyzing several different energy sources and paying attention to the relations energy-quality of life, energy-environment and energy-health.  Then will be discussed fossil fuels, renewable energies, the role of electric energy, nuclear energy, energy savings, the greenhouse effect, a possible sustainable development, the hydrogen economy and the risk and its perception. Then follow the conclusions and perspectives.

\end{abstract}

\end{frontmatter}
\section{Introduction}
\label{intro}

It is well known that the quality of life of the population has a direct relation with energy. Several indicators show the correlations between energy consumption and National Income procapite, energy and average lifetime, energy and infant mortality. With respect to the past, the work done by animals and slaves was replaced by machines that have become  more reliable and more efficient. The used energy sources changed quickly, passing from firewood, water mills, animal and slave work, to hydroelectric energy, coal, oil, gas, nuclear energy, solar energy, wind energy and others. The energy available at a low cost was a prime factor for improving the quality of life of the populations in the developed countries [1]: it is as if every citizen had a remarkable number of ``energy slaves'' [2]. The large energy consumption was a prerequirement to obtain more products and more services. The amount of products obtained per hour of work increased and the number of working hours per day decreased, leaving more time for leisure and culture.

The energy consumption had an enormous increase: in Italy in the years 1950-1970 the increase was about 10$\%$ per year per person, with an increase still larger for electric energy; currently the increase of the italian consumption is considerably inferior. 

Every energy source has, in greater or minor form, its ``difficulties''. As ``there is no rose without thorns'', there is no energy source without problems and pollutions: one may recall the use of firewood in early 1900s which led to the cutting of trees in large areas of the world, the sinking of oil tankers with recurrent wide scale pollutions, the nuclear incident of Chernobyl, the many dead miners in chinese coal mines. For the use of every energy source one should thus estimate the sanitary and environmental risks and study the methods for reducing them [3].
  
The energy issues occupy a prominent position in the economy and development, both at the national and at the world level. ``Any future politics will have to identify and make reasonable projections of the energy needs, from availability of primary energy resources and to provide a diversification of the energy sources. It should also insure that adequate scientific-technological potentialities be available to conjugate the well-being of the citizens and the preservation of the future habitat'' [2, 3]. 

After the 11 september 2001 it became obvious that it is also necessary to consider the effects of possible attacks to dams, fuel warehouses, nuclear and conventional plants and study precautions and safeguard measures. 

In the past century the energy problem seemed to mainly concern the developed countries, that were in an effective privileged position. From the beginning of the new millennium have become obvious the demands, the ``hunger for energy'' of the emergent countries, in particular China and India. All want more energy, and this places new economic, political and environmental problems [4].

 In the following a short general picture of the energy problem is made, with special reference to the European and Italian situations. 

\section{Primary energies}

\begin{figure}[h]
\begin{center}
{\centering\resizebox*{!}{4.7cm}{\includegraphics{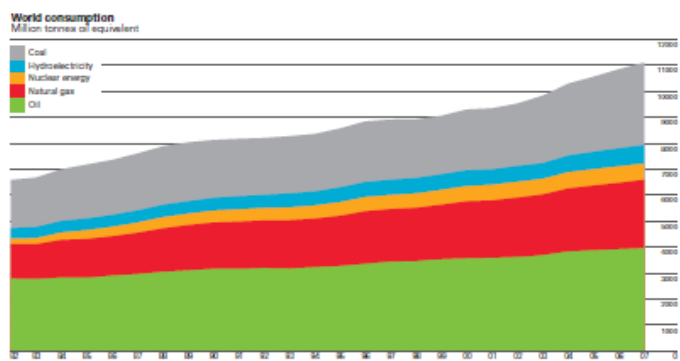}}\par}
\caption{\small World energy consumptions from 1982 till the end of the 2007 (BP, 2008) [5].}
 \label{fig:1}
\end{center}
\end{figure}

In 2003 the world consumption of energy was about 10 billion tons of oil equivalent ($\sim$10 Gtep), see Table 1 and Fig. \ref{fig:1} [5]; in 2007 the consumption was $\sim$11000 Gtep. The present average world increase of energy consumption is $\sim$3\% per year. The major part of energy comes from fossil fuels, in particular oil, Fig. \ref{fig:1}. But these fuels are not renewable and will eventually end; this is in particular true for oil. 

Italy does not have fossil fuels in sufficient quantity and depends from imports for $\sim$84$\%$ of its energy needs, a very large value. This causes problems of costs, balance of payments and of differentiation of sources. Moreover the Italian case is anomalous because we use percentages of  oil and gas larger than those of other nations, and because we have abandoned nuclear energy (a unique case).

We have also to consider the energy uses. In Italy, $\sim$35$\%$ of the energy is used by industry, $\sim$24$\%$ in the transports, $\sim$34$\%$ for civil uses and agriculture, $\sim$7$\%$ for non energy uses.

In the last decades there were many technological innovations that led to improvements of $efficiency$ $in$ $the$ $production$ (for ex. in thermoelectric plants which produce electric energy) and $in$ $the$ $use$ $of$ $energy$ (the luminous yield of electric lamps increased by more than a factor of 100 in a century, the yield of all engines improved, improvements in burning fuels for house heating). In the industrialized nations these improvements slowed down the increase of consumption and were like new energy sources [6].

\begin{table}
\begin{center}
{\small
\begin{tabular}{|c|c|c|c|}\hline
{    } & {Italy(2003)} & {EU-27(2004)} & {World(2003)} \\ \hline 
 Oil & 47$\%$ & 36.8$\%$ & 34$\%$ \\ \hline
 Gas & 33$\%$ & 24$\%$ & 21$\%$ \\ \hline
 Coal & 7.9$\%$ & 18.2$\%$ & 24$\%$ \\ \hline
 Renewables & 6.5$\%$ & 6.4$\%$ & 13$\%$ \\ \hline
 Nuclear & -- & 14.4$\%$ & 6.5$\%$ \\ \hline
 Imports of & & & \\ 
electric energy & 5.8$\%$ & -- & -- \\ \hline
 Total (100$\%$)(tep) & $\sim$193 Mtep & $\sim$1.17 Gtep & $\sim$10 Gtep \\ \hline
\end{tabular}}
\end {center}
\caption { Energy used annually in the world, in Europe (EU-27) and in Italy in tons of oil equivalent (tep) in 2003$/$2004 and percentages of primary energy sources (BP, ENEA). Procapite energy use in Italy in 2004: 3.1 tep$/$year$/$person, total CO2 emission from Italy in 2004: $\sim$451 Mt$/$year.}
\end{table}

\section{Fossil fuels}

The majority of the energy comes from $fossil$ $fuels$: their known world reserves at the end of 2007 are illustrated in Figs. 2, 3, 5 [5]. Notice that the largest reserves of oil are in the Middle East, those of gas in the Middle East and Eurasia, those of coal in Eurasia, Asia and North America. On the basis of present consumptions one may estimate that the oil may last $\sim$40 y, Fig. \ref{fig:3}, natural gas $\sim$60 y and coal $\sim$200 y (coal is the most abundant source and the reserves are well distributed on the planet). Notice that it is from 1982 that the oil companies say that the oil will last... 40 years, Fig. \ref{fig:3}! For the annual production of any fossil fuel we expect in the future a curve with a peak followed by a decrease. According to BP for the moment the decrease is not observed because technological developments allowed and still allow to discover new deposits at greater depths, under the sea, and to use deposits of lower quality {\footnote{In the 1930s one thought that  Lybia was only a ``bag of sand''. But when in the 1960s the technology allowed to reach oil deposits to depths of $\>$1500 m one could establish that a large amount of oil was present in Lybia.}}. According to the pessimists the oil peak will manifest in $\sim$2010, while according to the optimists the peak will manifest  in  2020-2030.

\begin{figure}[h]
\begin{center}
{\centering\resizebox*{!}{4.5cm}{\includegraphics{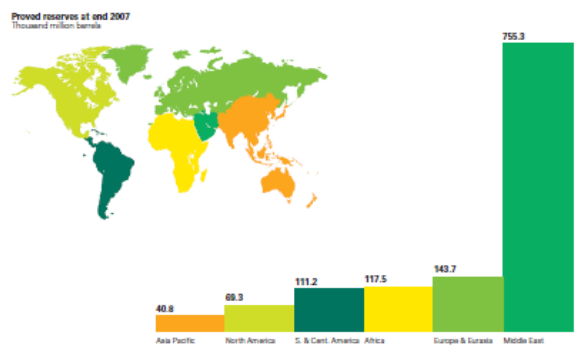}}\par}
\caption{\small  World reserves of oil at the end of 2007 (BP, 2008).}
\label{fig:2}
\end{center}
\end{figure}

In 2007 in Italy the extraction of oil was $\sim$5.9 Mt, and of natural gas $\sim$8.0 Mtep (the extraction of gas is decreasing; the reserves of gas in Italy are $\sim$270 billions m$^{3}$, situated mainly in the North Adriatic sea [http:$/$$/$www.enel.it$/$attivita$/$novita\_eventi$/$energy\_views$/$ archivio$/$2007\_01$/$schede\_tecnich$/$index.asp] (source ENEL and BP 2008 report). The oil and natural gas production ranks Italy in the 4$^{th}$ place of the European countries [7]. Even if limited, this production plays an important role. 

However the main problem is to find other sources of energy, which could be available for much longer times. It is possible that other fossil fields will be found in the world, in remote and cold zones, to greater depths and with higher extraction costs{\footnote{In the US and Canada there are large deposits of bituminous shales, in Canada also of oil sands. The increase of oil price made competitive the extraction of oil from the canadian oil sands (in the province of Alberta) and of heavy oil in Venezuela. ENI found oil sands in central Africa. Brazil found oil deposits at large depths in the Atlantic Ocean close to its coasts, and it seems that there are oil deposits around the Falkland islands  and close to the arctic coasts.}}. There will also be technological improvements in the use of these fuels, new energy sources will be found and therefore there will perhaps be a decrease in the fossils use. But there is a strong increase in energy consumption in the developing nations, and this is quickly modifying the situation. And the prices of the fossil fuels increased (the price of oil exceeded the 100 dollars per barrel and than decreased, see Fig. \ref{fig:4} and Section 11).

For the year 2030 is predicted a world population of $\sim$8 billion people and an annual world consumption of 17-18 Gtep per year [4]. In many forecasts these values are exceeded and... if proper alternative energy solutions are not found... then it will be necessary to burn everything, even... the furnitures of the houses!

\begin{figure}[h]
\centering
 {\centering\resizebox*{!}{6cm}{\includegraphics{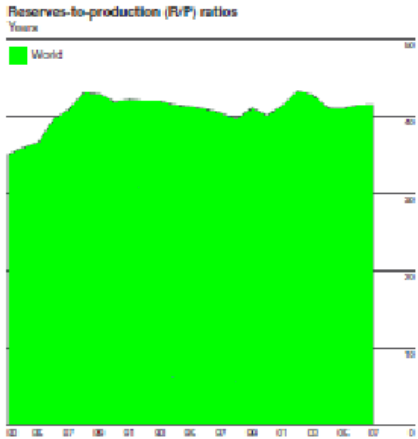}}}
\caption{\small The annual world ratio Reserve/Production (R$/$P) of oil versus year: notice that R$/$P remained at the level of 40 y  from 1982 to 2007 (BP, 2008).} 
\label{fig:3}
\end{figure}

Methods are being developed to use fossil fuels more efficiently and in a less polluting way, in particular for coal. The new coal plants are less polluting and more efficient than the older ones, see for ex. the new plant near Civitavecchia. There are also attempts to gasify coal producing ``syngas'' that can be used directly, or to produce liquid fuels and valuable chemical substances. These new methodologies are also benign for the environment ($Clean$ $coal$ $technology$).

\section{Renewable energy sources}

Renewable energy sources are shown schematically in Fig. \ref{fig:6}. Most sources are consequences of solar energy. A strong increase of these sources is expected, even if it is difficult that they can give an important contribution in the near future. The increase of the fossil fuel prices could bring a greater impacts to renewable energies (it also led to the search$/$exploitation of non conventional deposits of oil). But the recent economic and energy crises may lead to some recession and other consequences, see Sect. 11.

For the far future only two main energy sources are known: solar energy and the energy from nuclear fusion (for the midterm future there is also nuclear fission from U$^{238}$ and Th$^{232}$).

{\bf Solar energy} arrives everywhere, without cost, it is renewable and, on a small scale, does not pollute. It has the disadvantage of being diluted in space and to vary with the alternation day-night and of the seasons, Fig. \ref{fig:7}. The simplest applications are via solar panels, either for heating, for ex. the rooms in an apartment, or for the direct production of electricity [8]. The applications of small $photovoltaic$ $cells$ are  many, in little calculators and in wristwatches; larger panels serve to pump water from wells, power communication equipments and emergency systems, etc.

$Thermal$ $solar$ $panels$ reached economic competitivity. 

$Photovoltaic$ $solar$ $panels$ need technological improvements to increase the efficiency of the photovoltaic cells and reduce their production costs. There are many research activities aimed to reach these purposes. The $first$ $photovoltaic$ $generation$ is based on silicon solar cells (either monocrystalline or poli-crystalline). The $second$ $generation$ $of$ $photovoltaic$ $cells$ concerns thin film cells, polymeric cells, and many new technologies, for ex. with transparent varnishes on the glasses of the windows, etc. Remarkable developments are anticipated with the use of nanotechnologies (Quantum Dots) and thin multi-junction cells ( each covering a different region of the solar spectrum): some of these cells have reached  efficiencies of $\geq$40$\%$. ($Third$ $generation$ $solar$ $photovoltaic$) [9]. In any case there is a large increase of the solar industry.

\begin{figure}[t]
\centering
 {\centering\resizebox*{!}{5.5cm}{\includegraphics{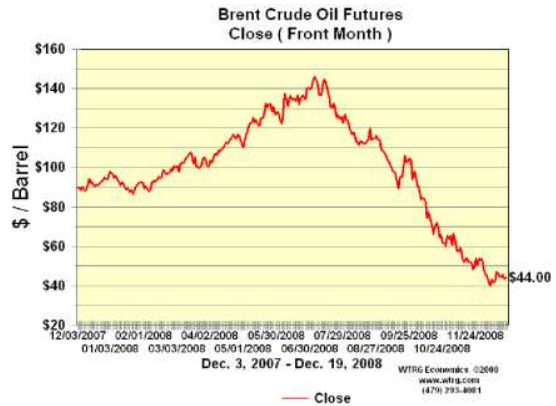}}}
\caption{\small Oil prices (in \$$/$barrel) in the last year.} 
\label{fig:4}
\end{figure}

Large surfaces are needed to obtain large amount of energy, with possible environmental consequences. There are incentives for the direct use of solar energy, in particular in Germany, Europe, United States, Japan: these projects contribute to standardize all components and reduce costs [8]. There are systems that use parabolic mirrors which focus the light on a thermal machine located at the focus of the mirror, Fig. \ref{fig:8}b; in other systems there are many mirrors that focus the light on a boiler at the top of a tower: high temperatures are reached, that may produce electric energy with higher efficiency, Fig. \ref{fig:8}a. Other systems use linear parabolic mirrors that focus the light on a line and high temperatures are obtained, Fig. \ref{fig:8}c [9].

\begin{figure}[h]
\begin{center}
{\centering\resizebox*{!}{4.5cm}{\includegraphics{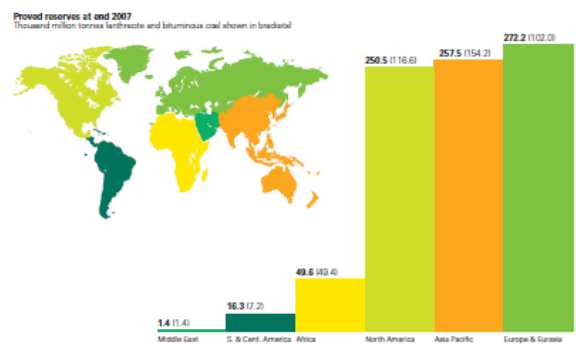}}\par}
\caption{\small World reserves of coal at the end of 2007 (BP, 2008).} 
\label{fig:5}
\end{center}
\end{figure}

The installation of small photovoltaic systems ($photovoltaic$ $roofs$) near the place of use reduces the transportation energy losses. Photovoltaic solar plants in the desert of Southern California can supply electric energy in the hours of top energy consumption, when air conditioners are turned on. In Australia a large ``solar tower wind plant'' is planned, where warm air, collected in a surface of many km$^{2}$, would raise in a large chimney setting in action many turbines with a total power $>$100 MWe, (the operating principle was verified with a tower in Spain). Here the investment cost  is high; the efficiency would be lower than that obtained in the systems illustrated in Fig. \ref{fig:8}; the cost of the KWh is still difficult to estimate. 

\begin{figure}[h]
\begin{center}
{\centering\resizebox*{!}{5.5cm}{\includegraphics{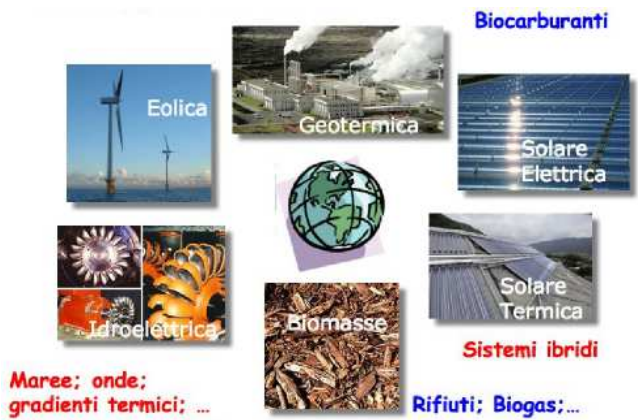}}\par}
\caption{\small Qualitative summary of renewable energies.} 
\label{fig:6}
\end{center}
\end{figure}

\begin{figure}[ht]
\begin{center}
\hspace{-0.5cm}
{\centering\resizebox*{!}{2.7cm}{\includegraphics{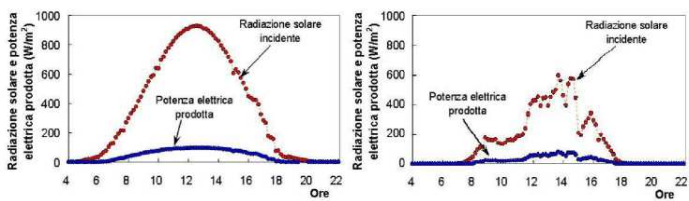}}\par}
\caption{\small Incident solar radiation and produced electric energy in a photovoltaic roof with solar plane panels. Every cell, made of silicon monocristals, has an efficiency of $\sim$18$\%$. The global efficiency of the system for the production of electric energy in a.c. current averaged over one year is $\sim$10$\%$.  
http:$//$www.scienzagiovane.unibo.it$/$pannelli.html.
} 
\label{fig:7}
\end{center}
\end{figure}

The intermittency of the solar source limits the maximum electric power which can be connected to an electric network: it is necessary to improve the network and to find systems which accumulate electric energy before connection to the net; further technological developments are needed.

 For the far future it was hypothesized the use of photovoltaic plants in orbit around the earth: there the day-night and seasonal variations do not exist. But how to transfer the energy to earth? It could be done, through microwaves. A test is planned with a low orbit satellite, sending microwaves to a small island of the Palau group in the Pacific.

 \begin{figure}[ht]
\centering
 {\centering\resizebox*{!}{3cm}{\includegraphics{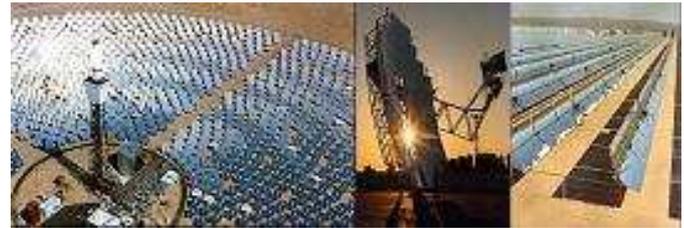}}}
\caption{\small Concentrated solar systems. For the system in the middle photo the efficiency for producing ac electricity is $\sim$30$\%$.} 
\label{fig:8}
\end{figure}

 {\bf Hydroelectric energy.} This was the first renewable source successfully used for electricity production on a large scale. Currently about 6$\%$ of the world energy comes from hydroelectric plants, Fig. \ref{fig:1}. In the developed nations, the large rivers are by now all used up with dams and hydroelectric plants. Only few small rivers remain available and one could only build small new hydroelectric plants. The situation is different in many developing nations, where there are unused rivers and their exploitation for hydroeletricity may bring economic progress. But one should remember that large artificial lakes and hydroelectric plants may bring environmental problems. 

There are hydroelectric plants with two water basins at two different levels: during the night one may pump the water to the high level lake and use the falling water during the day to produce electricity in the hours of top energy consumption.

{\bf Wind energy.} It is known and used since many years, above all in Holland, for flour mills and  to raise water from the ground. Wind energy comes from solar energy, which produces thermal imbalances in the atmosphere.
Electric energy can be obtained from wind turbines, with horizontal axes, placed at heights of $>$30 m to take advantage of more stable winds with less turbulence. Wind energy is becoming economically important in ``wind farms'' in zones with regular wind, like between the Atlantic Ocean and Europe: here the overall efficiencies are $\sim$20$\%$ for the direct production of electric energy. 

The European Union plans to obtain a relatively large fraction of electric energy from solar, wind and biomass plants {\footnote{In articles in newspapers there is often some confusion between $power$ and $energy$. Articles in newspapers quote often only the peak power of solar and wind plants, without considering their variability.}}.

{\bf Energy from sea tides.} The energy in the tides is large, but the fraction that can be used is small, $\sim$1-2$\%$ of the total. One needs high tides (of  $\sim$12 m), a re-entering coast to a river estuary that forms a natural funnel, etc. Since many years it is working the french plant in the Rance estuary, with a power of $\sim$240 MWe (in the moments of maximum and minimum tides). Many projects for tide plants have been made in England, Canada and USA. Around Italy the tides are small and it is nearly impossible to use them to produce energy.

{\bf Energy from sea waves.} In order to use the energy of sea waves one should make mechanical devices that absorb energy, reducing the amplitude of the waves. Small plants of 1-2 MWe of power are used in Japan and England. The released energy is fluctuating in time (it is zero when the sea is calm). To obtain the equivalent of one plant of 1000 MWe on the coasts of the Atlantic ocean it would be necessary to make installations with a total length of $\sim$400 km.

{\bf Energy from the sea currents.} The Gulf stream, from the Caribbean sea in central America to the coasts of northern Europe, has a total power of $\sim$26.000 MW on a front of hundreds of km. Also this energy comes from sunlight; one would need large areas for the installation of plants; but the energy taken could have consequences on the climate of Northern Europe. The current in the strait of Messina, in the sea arm between Calabria and Sicily, could be used. 

{\bf Energy from the difference of temperature between deep and superficial tropical seas.} One may use a vertical tube with a fluid which can be vaporized by the warm surface water and condensed by the deep cold water. The vapor could turn some turbines producing electricity. This exploitation is thinkable in tropical seas where there is a difference of $\sim$20$^{\circ}$C between surface and deep waters. Tests are being carried out in the Pacific ocean near the Hawaii islands. 

\begin{figure}[ht]
\centering
\hspace{-0.5cm}
 {\centering\resizebox*{!}{4cm}{\includegraphics{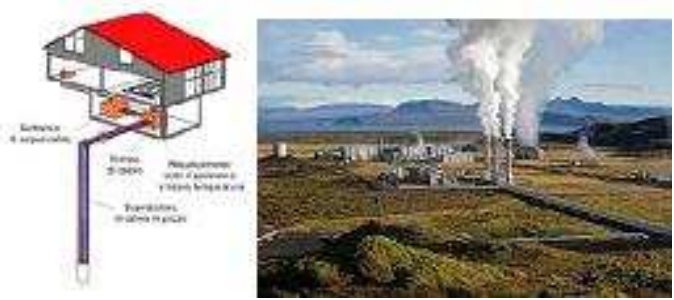}}}
\caption{\small (a) Scheme of a system  for domestic heating with a ground connected heat pump. (b) The geothermal plant of Nasjavellir in Island.} 
\label{fig:9}
\end{figure}

{\bf Biomasses.} (Fuel obtained from herbaceous and wood biomasses and municipal organic waste). Wood and other biomasses are still the most used fuels in many developing countries. The use of small biomass plants that use wood waste of every type can certainly be encouraged. 

{\bf Bio-fuels.} $Bioetanol$ and $biodiesel$ can be used as additives to benzine and diesel engines;  biomasses can also be used to produce biogas. Brazil obtains bio-fuels from sugar cane cultivations. Other nations hope to produce bio-fuels from various plants. But one must be sure that such productions do not enter in contrast with normal agriculture to produce food: probably the recent increases of the prices of grain, rice and corn are partially connected with the use of biomasses to produce bio-fuels. For the future one may produce biomasses from seaweeds. 

{\bf Energy from city wastes.} The re-use of city waste may serve to recover important raw materials, to produce fertilizers and bio-fuels. Moreover, through recycling the amount of waste diminishes. There are many plants that produce energy from city wastes. Energy may be obtained in heat form, burning the wastes (a high temperature is needed in order to destroy all complex molecules, such as dioxin) in incinerators, or through fermentation producing biogas. For the future it is possible to obtain a relatively important contribution of energy from city wastes. But the Napoli disaster indicates the importance of a proper programming.  

{\bf Geothermal energy.} Energy is obtained taking advantage of the inner heat of the earth, sees Fig. \ref{fig:9}. The geothermal gradient gives a measure of the temperature increase with increasing depth: it is $\sim$3$^{\circ}$C$/$100 m. In order to use geothermal energy one may use vapor systems, water at high temperature, warm water at lower temperature, and warm rocks. The first two methods are currently appropriate to produce energy from $natural$ $geysers$ and warm basins near the terrestrial surface in active zones, like Larderello in Tuscany, in Iceland, in the north-west USA, in the Philippines, Mexico, Japan and other countries; the third method uses $heat$ $pumps$ (Fig. \ref{fig:9}a). The fourth method (warm rocks), offers promising possibilities through rock fracture (in order to increase the surface on which heat exchange happens) and injection of cold water. Many systems for thermal uses have been installed [11]. In Italy it is in function the geothermal complex of Larderello, with $\sim$400 MWe of power.

{\bf Hybrid systems. Heat pumps.} These systems are not sources of energy, but they may optimize energy cycles and the exploitation of primary sources. The $heat$ $pumps$ work like refrigerators, but they do not remove heat from the system that one wants to cool; they remove heat from the external atmosphere and yield such heat to the system that one wants to heat (for ex. a house in winter, connecting the heat pump to external air or to an underground site, see Fig. \ref{fig:9}a). The system needs electric power; but globally there is an improvement in efficiency and thus in saving energy. Heat pumps are also used to cool houses in the summer [10].

\section{Electric energy}

\begin{figure}[ht]
\centering
 {\centering\resizebox*{!}{2.8cm}{\includegraphics{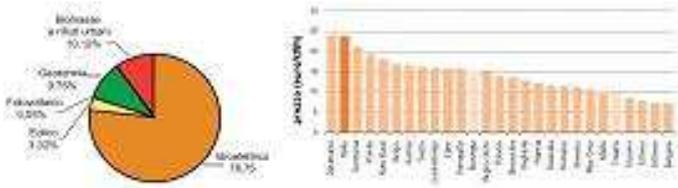}}}
\caption{\small (a) Production of electricity in Italy in 2004 per renewable source (in $\%$). (b) Prices (euro$/$Wh) of electric energy for domestic uses taxes included for annual consumption of 2500-5000 kWh$/$y.} 
\label{fig:10}
\end{figure}

$Electric$ $energy$ is an intermediate form of energy. It is produced in thermoelectric plants (burning oil, gas or coal), in hydroelectric and nuclear plants; smaller amounts are produced from the wind, photovoltaic and thermal plants; in Italy there are also the geothermal plants of Larderello. The conversion efficiency of fossil fuels into electric energy had several improvements: the new plants work at higher temperatures, are larger, less polluting and safer. Globally the efficiency is now $\geq$40$\%$: it can nearly double if the heat at lower temperature is used for heating buildings in winter and cooling them in summer. Fig. \ref{fig:10}a shows the electric energy production in Italy in 2004 from renewable sources (in $\%$): the larger contribution comes from hydroelectric plants, followed from geothermal plants and biomasses; smaller contributions come from wind and photovoltaic. In Fig. \ref{fig:10}b are indicated, for every european country, the prices of electric energy, including taxes, for domestic purposes, for annual consumptions between 2500 and 3500 KWh. In the second semester of 2007 the italian price (22,95 euro$/$KWh) was considerably higher than the european average price (14,20 euro$/$KWh) [12]. Also the price for italian industrial users was larger than the european average price. The high italian prices for electric energy are mainly due to the use of an expensive mix of fuels (mainly oil and gas); notice the large import of electric energy (see Table 1).

It is interesting to notice that the use of solar energy and energy from the wind, both subject to variability, are compatible with the use of plants optimized for continuity of production. 

Electric energy can be transported through integrated high voltage electric networks; after transportation the voltage is reduced and electric energy is used as mechanical, thermal, luminous, chemical energy. 

Electric energy plays a central role, perhaps more important than other forms of energy: it is versatile and its use is capillary in all fields {\footnote{I was in the middle of the New York black out in 1965: it seemed that the civil life had disappeared!}}. Electricity should be produced at the moment of use: one thus must face the general increase of the demand in the top hours (in Italy: from 9 to the 11 h and from 16 to 18 h). In these hours are turned on other plants, in particular gas and hydroelectric plants.
It is believed that the world consumption of electricity will double in the next 20 y.

Table 2 gives the mix of primary sources used to produce electric energy in 1973 and 2005 in the USA: notice that half of the electricity comes from coal (which is cheap); notice the increase of the nuclear source (mainly due to improvements in efficiency), the stability in the use of gas, the percentage decrease of hydroelectricity, the disappearance of oil  and the increase from renewable sources.
 
\begin{table}
\begin{center}
{\small
\begin{tabular}
{|c|c|c|}\hline
{\bf Fuel } & {\bf 1973} & {\bf 2005} \\ \hline
Coal & 45.5$\%$ & 49.7$\%$ \\ \hline
Nuclear & 4.5$\%$ & 19.3$\%$ \\ \hline
Gas & 18.3$\%$ & 18.7$\%$ \\ \hline
Hydroelectric & 14.8$\%$ & 6.5$\%$ \\ \hline
Oil & 16.9$\%$ & 3.0$\%$ \\ \hline
Others & 0.1$\%$ & 2.7$\%$ \\ \hline
\end{tabular}
}
\end {center}
\caption {\small Table 2. Mix of primary energies  used for the production of electric energy in the USA in 1973 and in 2005 [9] (In 2005 the production of electric energy in the USA was much higher than in 1973).} 
\label{table:3}
\end{table}

\section{Nuclear energy}

Nuclear energy is released in ``fission nuclear reactions" when a neutron breaks a heavy atomic nucleus in two or more lighter nuclei, produces two or more neutrons and yields energy in form of kinetic energy of the atomic nuclei and of the neutrons [13]. In the world are working 439 nuclear fission reactors that produce about 17$\%$ of the electric energy. In Europe (EU-27) there are 151 reactors that produce 43$\%$ of the electric energy [13]. France (59 reactors) and Japan (55 reactors) aim to produce their electric energy mainly with nuclear reactors and renewable sources (France achieved the aim (76$\%$ from nuclear), Japan is approaching it). The nuclear fuel is  U$^{235}$, which in natural uranium is less than 1$\%$. The cost of the nuclear  fuel is $\sim$15$\%$ of the cost of the electric kWh, while in an oil plant the cost of the fuel exceeds 50$\%$ of the final cost. The fission reactors under construction (2008) are 36 [7 in China, 6 in India, 7 in Russia, 3 in South Korea, 1 in Iran, ...]. 

\begin{figure}[ht]
\centering
 {\centering\resizebox*{!}{2.6cm}{\includegraphics{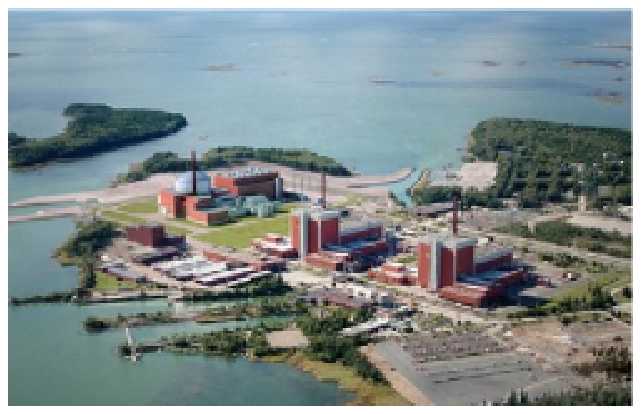}}}
 \hspace{-0.2cm}
 {\centering\resizebox*{!}{2.5cm}{\includegraphics{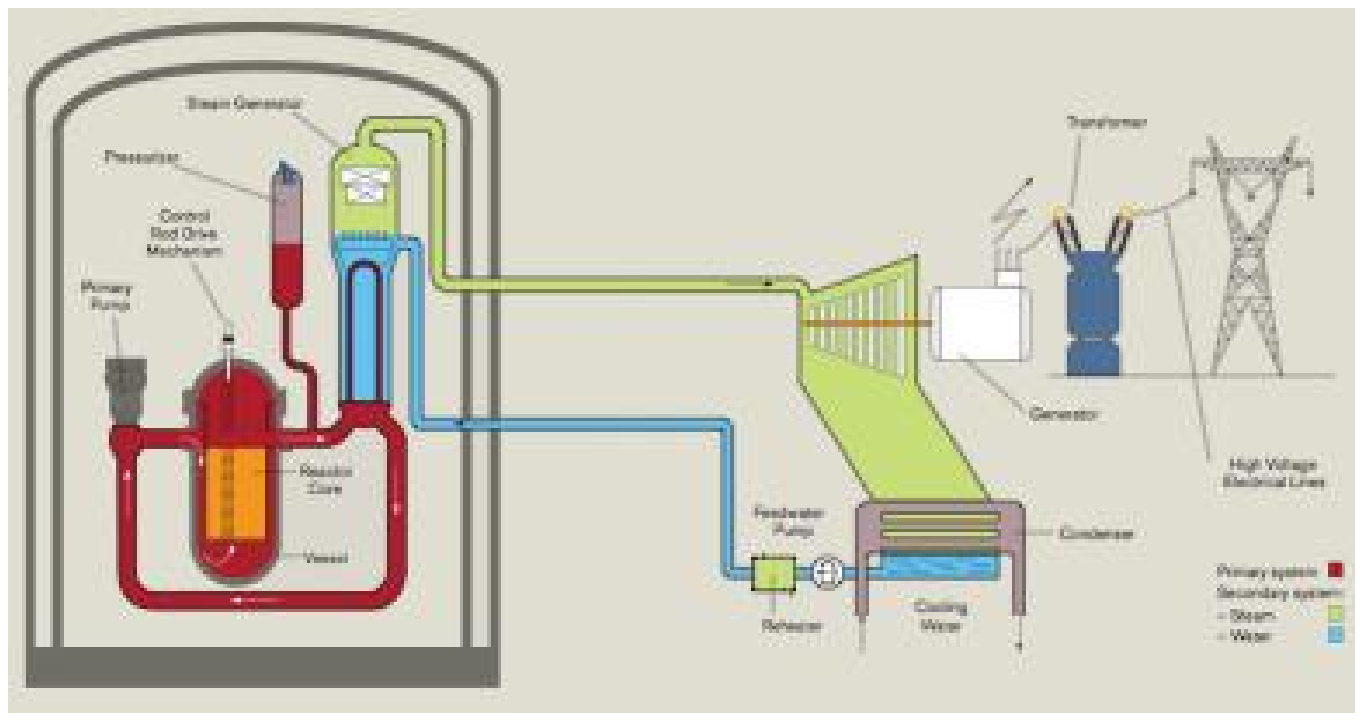}}\par}
\caption{\small (a) The nuclear finnish plants in Olkiluoto: on the right the 2 reactors already operational, on the left the new 3$^{rd}$  generation european 1600 MWe reactor with pressurized water (EPR). (b) Scheme of a water pressurized EPR.} 
\label{fig:11}
 \end{figure}

The nuclear reactors in use are $second$ $and$ $third$ $generation$ $reactors$; two reactors of the 3$^{rd}$ generation are in construction in France and Finland, others in China and India on French or US licenses, Fig. \ref{fig:11}. The future $4$$^{th}$ $generation$ $reactors$ are planned by an international consortium; they will be more efficient, will include $breeder$ $reactors$ and reactors for specific purposes. India and other nations are interested in Th$^{232}$ reactors {\footnote{ Th$^{232}$ is abundant in India. Th$^{232}$ needs partially enriched U$^{235}$ to start working in a reactor. Burning Th$^{232}$ does not produce long lived radioactive elements. With Th$^{232}$  is difficult if not impossible to make nuclear bombs.}}, for which further R$\&$D are needed. 

The U$^{235}$ reserves should not last for more than $\sim$80 years. Using fast neutron breeder reactors or reactors with accelerators it will be possible to use U$^{238}$ which is $\sim$100 times more abundant than U$^{235}$. Th$^{232}$ is three times more abundant than U$^{238}$. These fuels could be sufficient for thousands of years. 

The problem of long lived radioactive waste is important and many nations are investing in research $\&$ development. Various Countries, among which USA, France and Russia, developed reprocessing plants for nuclear fuels. Mixtures of oxides (MOX) are used as fuel in certain reactors: these produce energy, reduce long lived nuclear elements and produce some new fuel from U$^{238}$. There are many studies for fast breeder reactors and$/$or reactors with an accelerator which produce new fuel and eliminate long-lived waste. Russia uses a small fast reactor to burn plutonium bombs; a more powerful reactor in under construction {\footnote{ The USA-Russia agreement on ``Megatons to MegaWatts" allowed the ``burning" in nuclear reactors of thousands of nuclear bombs producing energy. At the end of 2007 a supplemental agreement concerned the burning of all plutonium bombs. Russia is burning them in a fast neutron reactor which also reduces long lived radioactive elements and breeds new nuclear fuel. Part of the money saved will go in projects in developing nations.}}. The reduction of long lived nuclear waste is part of the programs of the 4$^{th}$ generation reactors. The long lived nuclear waste may be destroyed with reactors with accelerator (ADS, Accelerator Driven Systems) that also produce energy (EA, Energy Amplifiers) [14].

One has to consider also the risks connected with the $proliferation$ $of$ $nuclear$ $bombs$. The reduction of this danger is connected with the signing of the non proliferation treaty by all nations which want to use nuclear reactors for energy production. It is important that more power is given to the IAEA for controls and inspections, [13]. The world nuclear future will depend mainly from oriental countries, Japan, China, India and South Korea, that have many programs in the field.

There is a lively discussion on the possible return to nuclear energy in Italy. Chicco Testa, an ancient opposer to nuclear energy, says that there are four reasons to return to nuclear energy: 1. The world wants more energy to exit from poverty. 2. it is very important to be free from fossil fuels; renewable energies are not yet alternative energies. 3. Nuclear energy does not produce greenhouse gases. 4. These gases are dangerous for world health. The ENEL company is engaging italian nuclear engineers for its nuclear plants abroad; it would be proper to have these ``qualified jobs" also in Italy; but it would be important that the choice to return to nuclear be ``bipartisan", in the wider sense of the term, in order to better find the economic resources needed and to reach a final decision avoiding any waste of money, like the waste made 20 years ago (in the exit from nuclear ``were thrown away" several tens of thousand billions of the old lire and was destroyed the human know out of physicists and nuclear engineers). The return could begin with one 3$^{rd}$ generation plant and with studies of 4$^{th}$ generation reactors. 

The other source of energy of the future is {\bf nuclear fusion}. The nuclear fusion reactions in the center of the sun transform four hydrogen nuclei in a helium nucleus, producing a large amount of energy. For the nuclear fusion reactors the fuel would be deuterium, that is heavy hydrogen (H$_{2}$, d) and tritium (H$_{3}$, t): d+t$\rightarrow$He$^{4}$+n. Deuterium is found in the sea water in a small percentage, but the total amount is large, sufficient for an indefinite period. Tritium, will be produced in the reactor ``blanket". Fusion reactors would produce electric energy, hydrogen gas from electrolysis of water or from ``cracking" of water molecules at high temperature, and heat at lower temperature. 

The $magnetic$ $confinement$ $fusion$ $reactor$ $ITER$ is a large reactor which will be realized in France from a world consortium of Europe, China, India, USA, Japan and South Korea. $Fusion$ $inertial$ $confinement$ $reactors$ with lasers are studied at the level of single nations. There still are many technical difficulties to plan and construct fusion reactors, and to make a final choice between magnetic confinement and inertial confinement. The time needed to establish feasibility is some tens of years {\footnote{ There are some studies about $could$ $fusion$, with results that are not too reproducible.}}.  

\section{Energy saving}

Several technological improvements allowed remarkable energy savings, but it is necessary to individuate savings that do not affect the standard of living  {\footnote{ At the University of Bologna, at the time of student unrests, at the end of an energy seminar a girl student made a brilliant intervention criticizing society and energy wastes. She was the last person to speak, went away looking satisfied, went to her car, a nice luxurious sport car, and left with a strong acceleration... she did not consider that her sport car and strong accelerations could be an energy waste, and that not only ``the others" should save energy, but also her.}}. 

{\bf Improvements in energy production.} 

$Natural$ $gas$ $from$ $oil$ $wells$. In the past, the gas from oil wells was burnt directly at the wells: from space one could easily spot these luminous sources. Now one uses most of this gas, transported through pipelines and ships, liquefied, and re-gasified. 

$Energy$ $Transportation$. Some energy is lost in the transportation from the place of production to that of use. For oil and coal one tries to optimize the routes and the sizes of ships. For electric energy the loss is reduced increasing the voltage of the transmission lines. 

\begin{figure}[ht]
\centering
 {\centering\resizebox*{!}{3.2cm}{\includegraphics{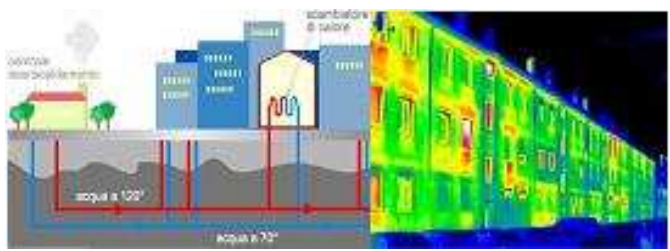}}}
\caption{\small (a) Scheme of teleheating. (b) Infrared photograph: in the clearer zones the buildings loose heat.} 
\label{fig:12}
\end{figure}

$Improvement$ $of$ $efficiencies$ $of$ $electrical$ $plants$: there was a continuous improvement: now we need to use the warm water produced by the plants to heat buildings, Fig. \ref{fig:12}a.

{\bf Improvements in the use of energy.}
 
$Improvements$ $of$ $the$ $efficiencies$. $Electric$ $Lamps$. In 1879 T. Edison made the first light bulb with a coal filament, which worked for $\sim$40 h. From that moment began a race to place lights in houses and in the roads. The efficiency of every source improved, with new efficient lamps of much longer duration and smaller consumption. 
 
$Turn$ $off$ $lamps$, $TV$, $recorders$, $and$ $their$ $little$ $warning$ $lights$ in the time in which they are not used. 
 
$Solar$ $buildings$. $Domestic$ $heating$. It is possible to make important  savings, even if it may take a long time. It is possible to improve thermal insulation, use double window glasses, insulate the warm water tubes, etc. For new buildings one may plan proper orientation to the sun, and appropriate use of heat pumps. Lowering a little the temperature inside buildings yields considerable saving; use the warm water from electrical plants, Fig. \ref{fig:12}a [6]; note the infrared photo in Fig. \ref{fig:12}b that allows to see where the buildings loose heat.
 
$Transports$. $Cars$. The car makers made improvements in aerodynamic shape, car weight and combustion. The drivers should avoid abrupt accelerations, slow down using the engine and reduce the cruising speed.
 
$Applications$ $of$ $solar$ $energy$ $in$ $agriculture$. $Warm$ $greenhouses$ may help to increase the yields and rationalize several cultivations. The use of low cost thermal solar panels (or warm water from plants) can supply the energy needed to dry agricultural products.

\section{Energy, environment and health. The greenhouse effect}

The use of any energy source has some impact on the environment. The earth temperature is determined from the energy balance between the incident solar radiation and that re-emitted towards space. Solar energy is maximal at wavelengths corresponding to visible light; a fraction of the incident energy is re-emitted in space. The atmosphere is transparent to visible light and absorbs the infrared radiation, which remains trapped in the atmosphere. If the atmosphere did not exist the average equilibrium temperature of the terrestrial surface would be $\sim$15$^{\circ}$C, while the measured value is $\sim$30$^{\circ}$C: this is the {\bf greenhouse effect}. In reality the processes that happen in the atmosphere are complicated. The gases that absorb the infrared radiation ($greenhouse$ $gas$) are  water vapor, carbon dioxide, ozone, methane and others. The greatest attention is concentrated in CO$_{2}$. From 1860 to 1995 the CO$_{2}$ concentration in the atmosphere increased from 300 parts per million (ppm) to 360 ppm. This may have lead to an increase of the air temperature of $\sim$1$^{\circ}$C, a fraction of which due to man. There are studies to place the CO$_{2}$ in deep geologic deposits. Many worries are expressed on human activities which enhance the greenhouse effect, even if many people think that the main source is of natural origin.

The populations which had greater energy consumptions had many improvements in their quality of life (but they now have more persons overweight). The heating of the houses caused the disappearance of some diseases and reduced the possibility that others became dangerous for their health. For what concerns the environment and health we recall air pollution from the cars and the traffic incidents. 

\section{Sustainable development. The hydrogen economy}

The notion of sustainable development appeared in a report of the United Nations in 1980 and became common $\sim$10 years later. The ``Declaration of Rio de Janeiro on environment and development'' of 1992 defines in 27 principles what ``sustainable development'' means; Principle n.1 says: ``Human beings are at the center of sustainable development. They have the right to a healthy and productive life in harmony with nature''. Such right ``must be realized so as to satisfy the requirements of environment and development of the present and future generations''. ``Agenda 21'' gives the priorities in 100 areas of intervention. The document is based on the principle of globality of the environment and on environment and development. 

The reduction of environmental effects from the energy cycles demands to increase their efficiencies, to privilege fuels low in carbon and renewable energies, improve energy saving, control the demographic factor. The UN discussed in many conferences the greenhouse effect and sustainable development. Limits on CO$_{2}$ emissions were fixed: it is difficult that Italy maintains its engagements and may have to pay heavy fines or acquire ``green bonds''. France will be ok because its nuclear plants were declared ok by the EU, since they do not emit CO$_{2}$. One should guarantee that future generations and emerging nations have a habitat with good living conditions [15].

By burning hydrogen in air one obtains water, which does not pollute; but there is no free hydrogen gas on earth. It was thought to use plants to produce electric energy and hydrogen gas (by electrolysis or by ``cracking water'' at high temperatures). In a ``$hydrogen$ $economy$'' the energy plants may be localized far away from cities and the produced hydrogen could be sent via gas lines; or hydrogen may be produced in small plants via photochemistry; but attention: hydrogen is a flammable gas! Technical improvements are needed to obtain a hydrogen economy. 

A $fuel$ $battery$ needs hydrogen or hydrocarbons and oxygen or air; it is not a primary source, but it is useful to reduce the emissions of greenhouse gas, and has a high efficiency for producing electricity [12]. 

\begin{figure}[ht]
\centering
 {\centering\resizebox*{!}{6cm}{\includegraphics{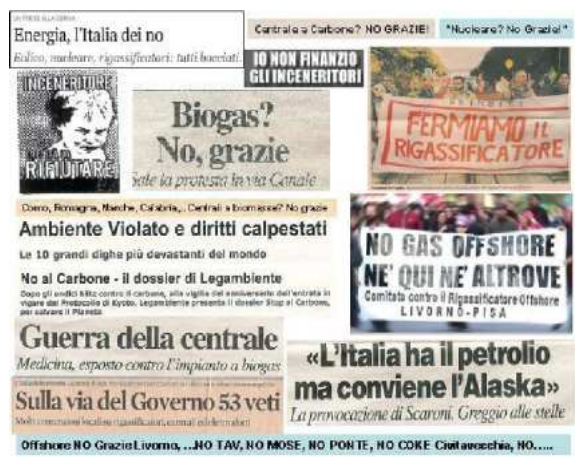}}}
\caption{\small A ``summary" of some ``difficulties" encountered in Italy: almost all are consequences of the so called $NIMBY$ $effect$, ``Not In My Back Yard".} 
\label{fig:13}
\end{figure}
\vspace{-0.4cm}

\section{Risk and its perception}

The risk evaluation is easily made when the frequency of the event that is the cause of the risk is high (for ex. car incidents); it is difficult for rare events, which may have catastrophic consequences (for ex. crash of two airplanes, incident in a nuclear plant, breach of a dam) [3]. This is the case for events with probabilities of $10^{-6} \div 10^{-7}$ per year. But the general attitude of the population is to estimate the risk instinctively. Often such {\it ``perception of risk''} is different from the real value. There is a tendency to underrate risks like smoke, car accidents, electricity, while other risks are overrated, for ex. those coming from any type of plant, in particular nuclear plants [16]. Often the ``$mass$ $media$'', that prefer sensationalism to ``the more grey'' correctness of information, contribute to such perception. It is difficult to convince the population to install any type of plant in their territory: such difficulty is generalized in Italy, Fig. \ref{fig:13}. One speaks of the $Nimby$ $effect$ (Not in my backyard) and may recall the difficulties to localize sites for garbage incinerators, biomass plants (in the Emilia-Romagna and Marche regions: sites approved at the regional and provincial levels, refused at local level), the protests for re-gasifiers. If we will not succeed to reduce such problems we will ``lose many trains" and could run the risks of  countries in decline. In order to reduce the oppositions in Fig. \ref{fig:13}, the italian government has alway given economic incentives to the regions and local communities which accept to have energy plants in their territory. 

To estimate the $environmental$ $and$ $sanitary$ $risks$, one has to search for the relevant parameters and the numerical values that give the limit of separation between an acceptable and an unacceptable risk. Efforts were made to determine the best environmental and sanitary limits, without forgetting the risk estimated on scientific bases and the perception of risk by the population.

\section{Conclusions. Perspectives.}

The Italian, European (and world) energy situations and the relations with health and environment were analyzed in many conferences and reports of National and International Institutions. The energy necessities should be faced with realistic programs, short term and long term. It does not exist a single form of energy that solves the energy problem; we must try to be ready to use all available energy sources and possibly ``invent new ones''. In more details, one needs to:
\begin{enumerate}
\item Develop national energy programs which are feasible and eco-sustained {\footnote{ At this moment the developing nations are mainly worried about energy and may neglect the environment: they will worry about it at a later time.}} . 
\item Upgrade all efficiency programs for any form of energy. An efficiency improvement is like a new energy source. 
\item Increase, upgrade and fund properly energy research. 
\item Use {\bf all forms of economically valid energies}, trying to solve and eliminate the objections illustrated in Fig. \ref{fig:13}. 
\item Reduce the dependence on fossil fuels. 
\end{enumerate}

These programs should be related to the demographic increase and the demands from the weaker part of the population to social improvements. They should be connected with the economic cost of each energy source and with long term programs. In Italy there are many research activities in the energy fields; but there is no real National Plan; probably little was planned and of the little programmed little was respected. We hope that the Italian ingenuity aids  us and ``saves us".

This lecture was prepared when the oil price was rising and seemed to keep rising forever. Then happened the financial crisis, the economic crisis, the slowdown of the economy and the strong reduction in oil prices (see Fig. \ref{fig:4}). How all this affects the energy problem? It is difficult to give an answer. At first sight it may slow down the search and deployment for new oil deposits, in particular at great depths, and of unconventional oil, see page 2, note 2. There may also be a reduction of investments in the most expensive renewable energies. But since all governments will try to decrease unemployment, there may be vast national programs to reduce unemployment, help renewable energies, in particular the solar energy field, and to provide extra investments in energy research. [17]. 

\section*{Acknowledgments} 

I thank many colleagues for suggestions and explanations, in particular V. Balzani, L. Bruzzi, F. Casali, D. Coiante and A. Lanza. I thank A. Casoni, M. Errico and M. Giorgini.


\begin{thebibliography}{9}

\bibitem{1} F. Casali, Energia pulita: quale?, Cappelli Ed., 1987.     \\ 
P. Angela e L. Pinna, La sfida del secolo, Mondadori Ed., 
ISBN 88-04-56071-1, 2006.          \\
N. Armaroli e V. Balzani, Energia oggi e domani: prospettive, sfide, speranze. 
Bononia Univ. Press, 2004.\\ 
Energia in Italia: problemi e prospettive (1990-2020), Study of the Italian 
Physical Society (SIF), 2008. 

\bibitem{2}  N. Armaroli e G. Balzani, Gli schiavi energetici, Rivista KOS, 
n. 243, dicembre 2005; http:$//$www.scienzagiovane.unibo.it$/$letture$/$KOS.pdf

\bibitem{3} Workshop ``L'uomo e l'ambiente: rischi e limiti di 
accettabilit\`a'', Sogesta, Urbino, Fondazione ENI Enrico Mattei, Pitagora 
Ed., 1994, ISBN 88-371-0692-0. In particular: \\
G. Campagnola e C. Palumbo, La produzione di energia dai rifiuti;   \\
R. Mazzucchelli, Tutela dell'ambiente e trasporto di idrocarburi;\\  
L. Bruzzi et al., Considerazioni conclusive.\\
S. Lombardini e R. Malamon, Rifiuti e ambiente, Il Mulino, ISBN 88-15-04191-5, 
1993.  

\bibitem{4} A. Clo, Energia e Tecnologia, Compositori, ISBN 88-7794-483-8 
2004.\\
A. Clo, Il rebus energetico. Tra politica, economia e ambiente, Il         
Mulino, ISBN 978-88-15-11514-0, 2008.

\bibitem{5} British Petroleum (BP), Statistical Review of World Energy 2007 
and 2008: http:$//$www.bp.com$/$ \\
US Department of Energy: Energy, sources and production:
http:$//$www.energy.gov$/$sources$/$index.html \\
ENEA, http:$//$www.enea.it, Annual Report 2008. \\
International Energy Agency (IEA):  http:$//$www.iea.org$/$.

\bibitem{6} L. Bruzzi e S. Verit\`a,  \\
http:$//$www.scienzagiovane.unibo.it$/$risparmio-energetico.html

\bibitem{7} http:$//$www.aspoitalia.it   \\                
http:$//$www.enel.it$/$attivita$/$novita\_eventi$/$energy\_views$/$archivio$/$ 

\bibitem{8} Solar panel: http:$//$it.wikipedia.org$/$wiki$/$Pannello solare\\
Solar panel:  http:$//$www.scienzagiovane.unibo.it$/$pannelli.html\\
Million roofs in USA: http:$//$www.eren.doe.gov$/$millionroofs

\bibitem{9} Photovoltaic Centre of Excellence, Australian Research Council, 
2007 report.\\
The centers Fraunhofer in Germany :   http:$//$www.ise.fhg.de$/$\\ 
D. Coiante, Il fotovoltaico di terza generazione : \\
http:$//$ricchezza-fotovoltaico.jujol.com$/$2008$/$02$/$23$/$il-fotovoltaico-di-terza-generazione-di-domenico-coiante$/$ 

\bibitem{10} http:$//$it.wikipedia.org$/$wiki$/$Pila\_a\_combustibile, \\
http:$//$it.wikipedia.org$/$wiki$/$Pompa di calore

\bibitem{11} M. H. Dickson, M. Fanelli, Cos'\`e l'energia geotermica: \\
http:$//$iga.igg.cnr.it$/$documenti$/$geo$/$Geothermal$\%$20Energy.it.pdf 

\bibitem{12} Autorit\`a per l'energia elettrica e il gas : Relazione Annuale 
2008, \\
http:$//$www.autorita.energia.it$/$relaz\_ann$/$relaz\_annuale.htm 

\bibitem{13} International Atomic Energy Agency, Nuclear Reactors : \\
http:$//$www.iaea.org.

\bibitem{14} Proposte di reattori nucleari con preacceleratore     \\
http:$//$en.wikipedia.org$/$wiki$/$Energy\_amplifier  \\
TRASCO: TRAsmutazione SCOrie:\\
http:$//$192.107.61.51$/$ADS$/$frame.html

\bibitem{15} L. Bruzzi et al., Sostenibilit\`a ambientale dei sistemi 
energetici.   \\
Tecnologie e Normative. Rapporto ENEA, ISBN 88-8286-164-8 (2007).

\bibitem{16} P. Volpe, Rischio reale e percepito, Sapere, Aprile 1996, pg. 44.
\\
G. Giacomelli e R. Giacomelli, No grazie! La sindrome Nimby in    
Italia, Natura e Montagna, Luglio-Ottobre 2008 pg. 64, ISSN 0028-0658. 

\bibitem{17} G. Giacomelli, "The Standard Model of Particle Physics. Neutrino Oscillations", Rad. Meas. 44 (2009) 826, arXiv:0901.2492 v2 [hep-ex]. 
\end{thebibliography}
\end{document}